\documentclass[a4paper]{JHEP3}

\usepackage[applemac]{inputenc}

\usepackage{amsmath}
\usepackage{amssymb}
\usepackage{subfig}
\usepackage[pdftex]{graphicx}

\title{Consistency of Relevant Cosmological Deformations
on all Scales}

\author{
Felix Berkhahn$^{ac}$, Dennis D.~Dietrich$^{b}$, and Stefan Hofmann$^{ac}$\\
\hskip -1.4mm$^a$Excellence Cluster Universe, Boltzmannstra{\ss}e 2, 85748 Garching, Germany\\
\hskip -1.4mm$^b$CP$^\mathit{3}$-Origins, Centre for Particle Physics Phenomenology, University of Southern Denmark, Campusvej 55, 5230 Odense M, Denmark\\
\hskip -1.4mm$^c$Arnold Sommerfeld Center for Theoretical Physics, Ludwig-Maximilians-Universit\"at, Theresienstra{\ss}e 37, 80333 Munich, Germany\\~\\
E-Mails: \email{felix.berkhahn@physik.lmu.de}, \email{dietrich@cp3.sdu.dk}, \email{stefan.hofmann@physik.lmu.de}
}
\abstract{Using cosmological perturbation theory
we show that the most relevant deformation of gravity 
is consistent at the linear level. 
In particular, we prove the absence of unitarity 
violating negative norm states in the weak coupling
regime from sub- to super-Hubble scales. 
This demonstrates that the recently proposed 
classical self-protection mechanism of
deformed gravity extends to the entire 
kinematical domain.    
}


\preprint{CP3-ORIGINS-2011-?\\LMU-ASC ?/11\\TUM-HEP-?/11}

\begin{document}

\maketitle


\section{Introduction and Outline}
Given the tremendous progress in high-precision cosmology,
in particular, the decisive character of distance indicators 
and structure formation probes on large scales,
the time is ripe to test the rigidity of Einstein's theory 
of gravitation on cosmological scales. 
This observational challenge is preceded by theoretical 
efforts aiming at consistent modifications of gravity
at the largest observable distances.
Obviously, only consistent theories are worthy to be 
confronted with data.
 
In a classical theory, at the exact level,
consistency refers to the existence of a well posed
initial value formulation and continuous solutions 
for the underlying degrees of freedom on the entire 
spacetime manifold. More precisely, at the technical level, the evolution
of a scalar degree of freedom $\Phi$ 
on a spacetime manifold $\mathcal{M}$, should be 
given by a quasilinear, diagonal, second order 
hyperbolic equation 
\begin{equation}
	q^{\mu\nu} \left(x;\Phi;\nabla\Phi\right)
	\nabla_\mu\nabla_\nu \Phi(x)
	=
	\mathcal{J}\left(x;R;\Phi;\nabla\Phi\right)
	\; ,
\end{equation}
where $q$ is a smooth Lorentz metric, which, in general is
not identical to the spacetime metric $g$, since it is
permitted to depend on the scalar degree of freedom
and its first derivative, and $\mathcal J$ is a 
smooth function that may have a nonlinear
dependence on these variables. Moreover, the current
density $\mathcal{J}$ may depend on the Ricci tensor $R(g)$. 

At the perturbative level, consistency of a classical
theory demands hyperbolic evolution only on a 
bounded spacetime region, the perturbative domain,
beyond which the fluctuation dynamics requires a 
non-perturbative completion that is consistent in 
the aforementioned sense. Perturbations around a 
classical solution can be quantized in the usual way,
given technically natural interactions. The standard 
requirements for a probabilistic interpretation offer yet another and distinct 
notion of consistency related to the 
quantum stability of the theory.

Classical stability at the perturbative level and quantum 
stability stand
on quite different footings. In fact, a finite domain 
of validity for classical perturbations does not cause any principal obstacle 
provided the underlying theory is consistent. Of course, once 
fluctuations leave the classical stability region their background
develops an instability towards a new ground state.  
In contrast, a quantum mechanical instability is not related to
specific initial conditions but instead to the massive production 
of particles at no cost, which are represented by negative norm states.
Therefore, the underlying theory is flawed at the fundamental level.
Additionally, what here is called quantum instability already has incisive effects within the framework of a purely classical analysis, which we discuss in Sec.~\ref{sec_class_inst}.

There are different frameworks for constructing consistent modifications of 
Einstein's theory of gravitation, once additional degrees of freedom are allowed
in the description\footnote{Hence, strictly speaking, these modifications
are not faithful deformations in the BRST terminology. This is known 
as the statement that multi-diffeomorphic theories have no Yang--Mills
analogue.}. 
As an instructive example\footnote{For a bi-diffeomorphic construction
see \cite{Damour:2002ws,ArkaniHamed:2002sp}.}, consider an additional second rank tensor
$\Psi$, not necessarily a metric, inducing the following relevant deformation
of the Einstein--Hilbert action
\begin{equation}
	\label{actFP}
  \mathcal{S}
	=
	\int d^4x \sqrt{-g}\;  M_{\rm P}^{\; 2}
	\left[R\left(g\right)-2\Lambda-m^2 H M H/2\right] 
	+
	\dots
	\; , 
\end{equation}
where $H\equiv g-\Psi$, $m$ has mass dimension one and sets the 
characteristic scale for the deformation, and $M(g)$ denotes the 
de Witt bimetric. 
Note that the de Witt bimetric is the most relevant albeit not 
unique choice for $M$, and we have neither written down explicitly 
the $\Psi$ kinetic and potential 
self-interaction terms nor the matter sector.

Assuming that $\Psi$ is locked into the Minkowski metric,
for one reason or another\footnote{For the moment it is not important to 
specify
 a dynamical mechanism that would give rise to the locking process.},
the interpretation of the deformation parameter follows from perturbing the 
metric around the Minkowski geometry, $g=\eta+h$. Expanding the action
(\ref{actFP}) to second order in the fluctuations $h$, the Fierz--Pauli theory \cite{Fierz:1939ix} is
rediscovered, for which the de Witt bimetric with respect to the background spacetime
is the unique unitary choice. 
This justifies to think of the deformation as a mass term with
the deformation parameter being the graviton mass. Of course, this interpretation 
hinges on the background geometry. 

The deformation presented in (\ref{actFP}) was primarily investigated on 
Minkowski and de Sitter background geometries for the following reasons:
Given the interpretation of the deformation parameter on a Minkowski background,
(\ref{actFP}) has been used to study consequences of a graviton mass for the 
principle of equivalence, in particular, how the impact of seemingly 
technically unnatural sources on the background geometry could be weakened. 
Higuchi \cite{Higuchi:1986py} showed that an intriguing relation between the deformation 
parameter and the cosmological constant needs to be fulfilled,
$m^2>1/3\Lambda\equiv H^2$ ($H$ stands for the Hubble constant), 
in order to render the free dynamics of $h$ on a de Sitter geometry unitary.
If this bound is violated, unitarity violating negative norm states are
introduced in the respective Hilbert space.

Both backgrounds are special in that no source specifications based on radiation or matter 
fields are required. This is of course different for generic Friedman cosmologies for 
which the Hubble parameter varies in time and, thus, the right-hand side of Higuchi's 
bound generalized to sourced Friedman geometries can be expected to become time dependent. 
In particular, it seems that for any deformation parameter at early enough times
unitarity violation is inevitable. The observation that the Hubble parameter's flow backwards in time
seems to induce quantum instabilities is a serious challenge for the viability of the
considered deformation. In fact, it is not clear whether the theory (\ref{actFP})
makes sense at all. 

In a recent paper \cite{Berkhahn:2010hc} we have already addressed the question of generalizing 
the Higuchi bound to generic Friedman spacetimes. This investigation relied on the usual
Stückelberg completion of $h$ in conjunction with the Goldstone boson equivalence
theorem \cite{ArkaniHamed:2002sp}. We found that the theory (\ref{actFP}) is, naively, 
subjected to two distinct bounds on Friedman cosmologies characterized by time dependent
Hubble parameters. One of them,
\begin{equation} 
	\label{unitarity_bound}
		m^2 >  H^2 +  \dot H
		\; ,
\end{equation}
enforces the absence of negative norm states (unitarity bound), whereas the second,
\begin{equation} 
	\label{stability_bound}
	m^2 >  H^2 + \dot H/3
	\; ,
\end{equation}
describes the region where hyperbolic evolution of the fluctuations 
is guaranteed (stability bound). Beyond this region, hyperbolicity breaks down. But this is 
no principal problem, since the breakdown is triggered by a strong coupling
regime that simply invalidates the perturbative approach, demanding for a 
nonlinear completion. Now, for all reasonable Friedman sources, $\dot{H}<0$.
As an important consequence, the (classical) stability bound imposes a stronger
requirement on the deformation parameter than the unitarity bound. 
For concreteness, we assign a value to the deformation parameter such that 
the stability bound is satisfied for times $t>t_*$. Evolving backwards in time,
the (classical) stability bound will eventually be invalidated since the 
deformation parameter is constant for the most relevant deformation
(\ref{actFP}). This signals the onset of the nonlinear regime. Thus, 
the would be unitarity bound lies beyond the perturbative domain and its 
derivation using perturbation theory cannot be trusted. In this precise sense,
the theory is self-protected against unitarity violations, and moreover, 
there is an open window of opportunity for a consistent nonlinear completion.  

Even though the Goldstone boson equivalence theorem represents a powerful
diagnostic tool that allows to extract the leading short-distance behavior
(and, futhermore, many interesting phenomena related to the most relevant deformation
of the Einstein--Hilbert term can be understood by employing it, as for example
the structure of the Fierz--Pauli mass term, the vDVZ discontinuity 
\cite{vanDam:1970vg, Zakharov:1970cc} or the Vainshtein radius 
\cite{Vainshtein:1972sx}, see also \cite{ArkaniHamed:2002sp}), 
it applies only in normal neighborhoods characterized by 
sub-Hubble distances $\ll 1/m$. 

The main purpose of the present paper is to extend our consistency analysis to
the intermediate and low energy regime. The prime framework to achieve this
is a full-fledged cosmological perturbation theory for all degrees of 
freedom. As usual, the metric fluctuations are decomposed into irreducible 
SO(3) tensors in accordance with the isometries of Friedman geometries.  
Compared to the $m=0$ case, the equation of motion for the 
second rank SO(3) tensor modes is deformed only by an additional hard mass
term. This is due to the fact that the degrees of freedom carried by the 
second rank SO(3) tensor are gauge invariant in the undeformed theory.
The equations of motion for the first (vector) and zeroth (scalar) rank SO(3) tensors
change considerably in the deformed theory. This is a testimony of the fact that
the deformed theory (\ref{actFP}) apparently has no gauge redundancy. It should be noted, however,
that the deformed theory has an equal amount of constraints compared to the 
gauge freedom possessed by the undeformed theory (and in fact could be understood
as the gauge fixed version of the Stückelberg extended theory).

The importance of these efforts is easily illustrated by the following results:
From the SO(3) vector sector arises a stability criterion that cannot be recognized
by employing the Goldstone boson equivalence theorem. This additional criterion
signals the presence of a tachyonic instability whenever
\begin{equation} 
{\bf k}^{\; 2}_{\rm phys} 
+ 3 \dot{H} + 2 m^2 \ge 0 
\end{equation}
is not satisfied. Here, ${\bf k}_{\rm phys}\equiv {\bf k}/a(t)$
denotes the physical wavenumber. On sub-Hubble scales, this
criterion is always fulfilled and, thus, the dynamics extracted by employing 
the Goldstone boson equivalence is not affected by the tachyonic instability
in the vector sector. In fact, the equivalence theorem does not cover this sector
at all, as it is subdominant compared to the scalar sector.  
In order to preserve stability on super-Hubble scales, however, we find the new bound 
\begin{equation} 
	\label{vector_bound_k0_intro}
		m^2 > - 3/2 \, \dot H \; .
\end{equation}
For any choice of the deformation parameter, this bound will be violated in 
the sufficiently early Universe, and, as a consequence, the vector modes
will develop a tachyonic instability, thereby triggering the transition
to a new ground state. This result supports the self-protection mechanism
found and analyzed in \cite{Berkhahn:2010hc}. The vector sector, thus, 
plays an important part in the stability analysis, although it does not 
participate in the Goldstone boson equivalence. 

The cosmological perturbation theory of (\ref{actFP}) reveals more insight 
into the stability dynamics, even in the scalar sector. Most importantly,
the unitarity bound (\ref{unitarity_bound}) seems at work on all scales and
not just on extreme sub-Hubble scales. Isolating the scalar sector, 
this poses a potential threat for the self-protection mechanism, since it is a priori not clear whether a strong coupling regime self-protects the theory also on super-Hubble scales. We have, however, shown analytically that the scalar sector is protected against unitarity violations for $\bf{k} = 0$ in the same sense as it was for sub-Hubble domains. To be more precise, we again find a stability violating region that occurs before the system enters the would-be unitarity violating region when evolved backwards in time. Compared to the sub-Hubble case, this region is simply shifted to larger values of the time $t$, so it seems reasonable to assume that there exists such a stability violating region for all values of $\bf k$. This conjecture is also confirmed by a numerical analysis. Moreover, as we have discussed, we know that the vector sector will become unstable whenever (\ref{vector_bound_k0_intro}) is violated, and thus contributes importantly to the self-protection of the system.

\section{The evolution of small fluctuations in the deformed theory}

The deformed equations of motion for the metric field $g$ following from
(\ref{actFP}) are given by
\begin{equation}
	\label{eom_fundamental}
	G_{\mu\nu}(g) - m^2
	M^{\; \; \alpha \beta}_{\mu\nu}(g) \; H_{\alpha\beta}
	= - 8\pi M_{\rm P}^{\; -2} \; T_{\mu\nu}(g,\chi)
	\;,
\end{equation}
where again $H=g-\Psi$. $\Psi$ is assumed to be locked into 
some reference metric, by one mechanism or another. $T$ denotes 
the energy-momentum source, which depends on matter and radiation
fields $\chi$, the metric field, and, in principle, 
an effective cosmological constant, as well. 
Any solution of the undeformed Einstein equations will be respected
by the deformation, provided $\Psi$ is locked into the appropriate 
tensor.

The Bianchi identity of the undeformed theory together with 
energy-momentum conservation of the source 
implies the following four exact constraints 
on the combination $H=g-\Psi$ in the deformed theory,
\begin{equation}
	\label{FPcon}
	\nabla^{\mu} H_{\mu \nu} - \nabla_{\nu} H 
	=
	0
	\; .
\end{equation}

Consider now metric perturbations $h=g-\gamma$ around a Friedman 
background $\gamma$ compatible with $T$. Assume $\Psi$ to be locked into
the respective Friedman metric and to be inert to the extend 
that it can be considered a fixed reference metric. Then $H=h$ and 
the equations of
motion for small $h$-fluctuations following from (\ref{eom_fundamental}) 
are 
\begin{equation} 
	\label{eom2}
	\delta R_{\mu \nu}(\gamma, h) 
	- m^2 \left( h_{\mu \nu} 
	+ \frac{1}{2} h \; \gamma_{\mu \nu} \right)
	= 
	-8\pi M_{\rm P}^{\; -2} \; M^{\; \; \alpha \beta}_{\mu\nu}(\gamma) \; \delta T_{\alpha\beta}
	\; ,
\end{equation}
to linear order in $h$. Here, $\delta R$ and $\delta T$ are 
the linearized Ricci and energy-momentum tensors, respectively. 
To this order, the four constraints \label{FPcon} are given by
\begin{equation}
	\label{effgauge}
	\nabla^{\mu}(\gamma) h_{\mu \nu} - 
	\nabla_{\nu}(\gamma) \gamma^{\mu\nu} h_{\mu\nu}
	=
	0
	\; ,
\end{equation}
which looks like a gauge constraint, but in fact is not.

The spatial isotropy and homogeneity of Friedman backgrounds
allow us to decompose the metric fluctuation $h$ into 
irreducible tensors with respect to these isometries,
\begin{eqnarray}
	h_{00} & = & -E \label{decompose_h00} \; ,\\
	h_{i0} &=& a\left[ \partial_i F + G_i \right] \label{decompose_hi0} \; ,\\
	h_{ij} &=& a^2 \left[ A \delta_{ij} + \partial_i \partial_j B + 
		\partial_{(j} C_{i)} + D_{ij} \right] \label{decompose_hij} \; .
\end{eqnarray}
Here, $E$, $F$, $A$, and $B$ denote SO(3) scalars, $G_i$ and $C_i$
are the components of a transverse SO(3) vectors ($\partial^a G_a=0, \partial^b C_b=0$),
and the $D_{ij}$ denote the components of a transverse-traceless rank-2 SO(3) tensor 
($\partial^a D_{ab}=0 $ and $\delta^{ab} D_{ab} = 0$).

The appropriate source for a Friedman spacetime 
is the energy-momentum tensor of a perfect fluid. Its perturbations can be decomposed
in the same spirit
\begin{eqnarray}
	\delta T_{00} &=& \delta \rho - \bar{\rho} \, h_{00}  
	\label {EMT_00} \; ,\\ 
	\delta T_{0i} 
	&=& 
	- \left( \bar{\rho} + \bar{p} \right) \delta u_i + \bar{p} \, h_{0i} 
	\label{EMT_0i} \; ,\\
	\delta T_{ij} 
	&=& 
	\bar{p} \, h_{0i} + a^2 \delta_{ij}  \delta p 
	\; ,\label{EMT_ij}
\end{eqnarray}
where the normalization condition $g(u,u) = -1$
and the background equation $\bar{u}^{\mu} = \delta^{\mu}_0 $ have been used. 
The three-velocity field ${\bf \delta u}$ will be decomposed in a gradient and a curl,
$\delta u_a = \partial_a \delta u + \delta u_a^V$.

Using the irreducible SO(3) tensors from (\ref{decompose_h00}-\ref{decompose_hij}),
the constraint (\ref{effgauge}) can be decomposed accordingly,
\begin{eqnarray}
	- 
	3 \dot{A} - \dot{\tilde{B} } + (\Delta/a^2) a F + 3 H E - 3 H A - H \tilde{B} 
	&=& 0 \label{effgaugeS1} \;, \\
	\partial_j \left[ - (aF \dot{)} - 3 H (aF) \right] - 
	\partial_j \left[ E + 2 A \right] 
	&=& 0 \label{effgaugeS2} \;, \\
	-(aG_j\dot{)} + \Delta C_j  - 3 H (aG_j) 
	&=& 0 \label{effgaugeV} \; ,
\end{eqnarray}
where $\tilde{B} \equiv \Delta B$. The constraint 
(\ref{effgaugeS1}) is obtained from the $\nu=0$ part of (\ref{effgauge}), (\ref{effgaugeS2}) from its $\nu=i$ part proportional to a gradient of a scalar, and (\ref{effgaugeV}) from its $\nu=i$ 
part given by a transverse vector.

Now, we have all ingredients to linearize Eq.~(\ref{eom2}) and to 
equate the rank-2,1,0 SO(3) tensor contributions separately. 

\subsection{Rank-2 contribution}

The rank-2 SO(3) tensor contribution results from the 
transverse-traceless part of the spatial-spatial 
components of (\ref{eom2}), and is given by
\begin{equation} 
	\label{eom_tensor_mode}
	 - \ddot{D}_{ij} -3 H \dot{D}_{ij} + \left(\Delta/a^2\right) D_{ij} -  m^2 D_{ij} = 0
	\;.
\end{equation}
It is worth mentioning that (\ref{eom_tensor_mode}) reduces to its counterpart
in the undeformed theory in the $m\rightarrow 0$ limit. This is a manifestation
of the fact that the constraint (\ref{effgauge}) cannot support 
transverse-traceless modes and, as a result, general relativity can be
continuously recovered in this sector. 
Provided the deformation parameter is small, $m^2 \lesssim H^2$,
the deformation term in (\ref{eom_tensor_mode}) will not change the
dynamics very much. In particular, the frozen mode on super-Hubble scales,
$-\Delta/a^2 \ll H^2 $,
is still present like in the undeformed theory.

Concerning stability, the equation of motion (\ref{eom_tensor_mode}) always yields
stable solutions, since the coefficients of both, the $D_{ij} $ and $\dot{D}_{ij} $ 
terms coincide with the sign of the coefficient in front of $\ddot{D}_{ij} $.
As a consequence, displacements will always be pulled back to the equilibrium position. 

In the following, we will always use the same symbol for both the real space and Fourier 
space amplitudes of any dynamical variable like $D_{ij} $.

\subsection{Rank-1 contribution}

The deformed equations of motion (\ref{eom2}) contribute two equations in the 
SO(3) vector sector of the theory, one from equating the spatial-temporal components,
the other from equating the spatial-spatial components. As we will see,
it suffices to consider the spatial-temporal equation together with the constraint
(\ref{effgaugeV}) and momentum conservation to solve the vector sector. 
The vector part of the spatial-temporal equation is given by
\begin{equation} 
	\label{i0_vec}
	16\pi M_{\rm P}^{\; -2} \left( \bar{\rho} + \bar{p} \right)  {\bf\delta u^V}/a 
	= 
	\left(\Delta/a^2  -  2 m^2 \right) {\bf G} -  (\Delta/a^2) a\dot{\bf C} 
	\; .
\end{equation}
For convenience, let us define $\widetilde{G}_j \equiv a G_j$.
From the constraints (\ref{effgaugeV}), it then follows that
\begin{equation} 
	\label{constraint_Cj}
	\Delta \dot{\bf C} 
	= 
	\ddot{\widetilde{\bf G}} + 
 	\Big(3 H \widetilde{\bf G}\dot{\Big)} \; .
\end{equation}
Inserting this equation into (\ref{i0_vec}) yields
\begin{equation} 
	\label{eom_G}
	16\pi M_{\rm P}^{\; -2} \left( \bar{\rho} + \bar{p} \right)  {\bf\delta u^V} 
	= 
	 - 
	\ddot{\widetilde{\bf G}} - 
	\Big(3 H \widetilde{\bf G}\dot{\Big)}
	- 	2 m^2 \widetilde{\bf G} + (\Delta/a^2) \widetilde{\bf G} \; .
\end{equation}
A solution for the divergence-free part or the three-velocity field
${\bf \delta u^V}$ can be obtained from the momentum conservation statement
in the corresponding sector, which is given by
\begin{equation} 
	\label{decay_uj}
	\Big( \left(\bar{\rho} + 
	\bar{p} \right) {\bf \delta u^V} \dot{\Big)} + 3 H \left( \bar{\rho} + \
	\bar{p} \right) {\bf \delta u^V} = 0
	\; .
\end{equation}
This shows that the quantity $(\bar{\rho} + \bar{p}){\bf\delta u^V}\propto 1/a^3$
decays and can therefore be neglected at late times. 
As a consequence, the equation of motion for ${\bf G}$ (\ref{eom_G}) 
is source-free at late times. 

Investigating the stability of (\ref{eom_G}), we see that the Hubble-friction
enters with the correct sign, whereas the terms with no time derivatives on ${\bf G}$
need to satisfy
\begin{equation} 
	\label{vector_bound}
	\left[-\left(\Delta/a^2\right) + 3\dot{H} + 2 m^2\right] {\bf G} \ge 0
\end{equation} 
to give a stable solution for {\bf G}. 
Surely, in certain kinematical regions and for particular 
values of the deformation parameter, the bound
(\ref{vector_bound}) will be violated, and, as a consequence,
a tachyonic instability will be generated.  
Indeed, for sufficiently early times, there will be such an
instability for all three-momenta, provided that 
$\dot{H}$ increases faster than $-\Delta/a^2$ for decreasing $t$.
This is the case, for instance, during radiation and matter 
domination, but not for the epoch when the cosmological constant
dominates. In the latter case, the vector modes are always stable
for arbitrary three-momenta.

On extreme super-Hubble scales, $-\Delta \ll (a H)^2$, 
the system develops instabilities whenever the bound
\begin{equation}
	\label{vector_bound_k0}
	m^2 \ge - 3/2 \, \dot{H}  
	\; .  
\end{equation}
is violated.
This bound is a new result that has not been obtained 
in the previous work \cite{Berkhahn:2010hc} based on
the Goldstone boson equivalence. The bound
(\ref{vector_bound_k0}) is instrumental for the self-defense
of the theory against unitarity violations: Consider an equation of state of the form
$p(\rho) = w \rho \;, w=$const. For $w<10/3$, the bound
(\ref{vector_bound_k0}) is even stronger than (\ref{stability_bound})
and, furthermore, supports the self-protection mechanism
described in \cite{Berkhahn:2010hc}. 

Once the equation of motion (\ref{eom_G}) for $\widetilde{\bf G}$
is solved, the constraint (\ref{effgaugeV}) allows to solve for
${\bf C}$ up to a spatially homogeneous contribution which, anyhow,
does not contribute to the spatial-spatial components of the 
metric perturbation, since ${\bf C}$ enters only with spatial
derivatives. This clearly shows that the vector sector contains
exactly one independent divergence-free three-vector field,
and, thus, is inhabited by two independent degrees of freedom.  

\subsection{Rank-0 contribution}

Like in the undeformed theory, the scalar sector is the most
intricate. It contains as geometric ingredients 
the scalars $A$, $B$, $E$ as well as $F$, and from the source $\delta \rho$,
$\delta p$, and $\delta u$. Not all of these variables
are, however, independent. Indeed, assuming a source with equation of
state $p = p(\rho)$ allows to reduce the dynamics to a set
of two coupled second-order differential equations 
for $A$ and $\widetilde{B}=\Delta B$:
\begin{eqnarray}
	\label{box_eom_A}
	\ddot{A}
	&=&
	- 3(1-w) H \dot{A} + w \left(\Delta/a^2\right)A
	- \left[2 m^2 - 6w\left(H^2-m^2/2\right)\right]A+
	\nonumber \\
	&& + w H \dot{\widetilde{B}}
	   + 2w\left(H^2-m^2/2\right) \widetilde{B}+
	\nonumber\\
	&& + H \dot{E}
	-m^2 E(A,B)
	\; ,\nonumber \\
	\\
	\label{box_eom_B}
	\ddot{\widetilde{B}}
	&=&
	- 7 H \dot{\widetilde{B}} - 4 \left(H^2+m^2/2\right) \widetilde{B} +
	\nonumber \\
	&& -12 H \dot{A} -3\left(\Delta/a^2\right) A - 12 H^2 A +
	\nonumber \\
	&& + \left(12 H^2- \Delta/a^2\right) E(A,B) 
	\; ,
\end{eqnarray}	
where $E$ is expressed in terms of $A$ and $B$,
\begin{eqnarray}	
	\label{box_E}
	&& \left[\dot{H} + \left(2-3w\right) H^2 - m^2\right]E(A,B) 
	= 
	\nonumber \\
	&& - \left(w-1/3\right) H \dot{A} - (w-1/3) \left(\Delta/a^2\right) A  
	- \left[\dot{H}+\left(1+6w\right) H^2-\left(2+3w\right) m^2\right] A +
	\nonumber \\
	&& - \left(w-1/3\right)H\dot{\widetilde{B}}
	-
	(1/3)\left[\dot{H}+\left(1+6w\right) H^2-\left(2+3w\right) m^2\right] \widetilde{B}
	\; .
\end{eqnarray}
The remaining geometrical SO(3) scalar $F$ can be obtained using
the deformation constraint (\ref{effgaugeS1}). 
Then $\delta \rho$ can be derived from the temporal-temporal component 
of the linearized deformed equations of motion (\ref{eom2}), and 
$\delta u$ can be derived from the spatial-temporal components of (\ref{eom2})
by extracting the spatial gradient contributions. Finally, $\delta p$
follows from the equation of state $\delta p = c_{s}^{\; 2} \delta \rho$
where $c_s$ denotes the isentropic sound speed in the source.
The details of this calculation can be found in the appendix.

\section{Stability analysis in the scalar sector}

In \cite{Berkhahn:2010hc} we have already discussed some qualitative differences between the two bounds (\ref{unitarity_bound}) and (\ref{stability_bound}): The former leads to negative norm states, which spoils the probabilistic interpretation of the theory, while the latter signals the breakdown of perturbation theory. In Sec.~\ref{sec_class_inst} we reiterate on the issue by presenting further arguments for the physical difference of both bounds, based purely on the classical evolution. After Sec.~\ref{sec_class_inst} we continue with the stability analysis in the scalar sector.

\subsection{Classical effects of the different types of instabilities}
\label{sec_class_inst}

As already mentioned above, what here is called quantum instability (that is the appearance of negative norm states in the quantized theory) already has an incisive effect within the framework of a purely classical analysis: Let us have a look at a setup, which is actually capable of capturing all the relevant physics at the linear level for sub-Hubble scales \cite{Berkhahn:2010hc}, based on the classical equation of motion for a scalar $\phi$, $\alpha\ddot\phi+  \epsilon \dot \phi + \beta\phi=0$. Here, the coefficients $\alpha$, $\beta$, and $\epsilon$ are functions of time. For $\alpha,\beta, \epsilon>0$ the system is stable. The classical stability bound manifests itself in a change of the sign of $\beta$ while $\alpha$ is still positive, which triggers an exponential instability, and the perturbative analysis breaks down. For a gradual zero-crossing the spring constant is already small before the hard bound is hit and the oscillations might enter the nonlinear regime already before the exponential instability is triggered. 

Nevertheless, we can still choose initial conditions that allow us to evolve the system for a small amount of time inside the region $\beta < 0$ until the fluctuation grows large. We can, however, not use this approach to try to cross the point where $\alpha$ turns negative as well, as close to this point, $\alpha$ is already small, and the effective spring constant has an extremely negative value, which goes to $-\infty$ just at the zero-crossing. Hence, in its vicinity, the time for which we can evolve the system in the just described fashion goes to zero. As a consequence, there is no reason why a change of sign of $\alpha$ {\sl after} a change of sign of $\beta$ should have any physical relevance for the full system. This is a manifestation of the self-protection mechanism.

Let us now consider the opposite case when $\alpha$ changes its sign before $\beta$ does. In this case, the effective spring constant $\beta/\alpha$ grows big before the zero-crossing of $\alpha$, confining the oscillations of $\phi$ to small values even more. The equation of motion, however, runs into a singularity because the term with two time derivatives (thus terminating the time evolution of the system) vanishes. Hence, this case would be much more severe, as the system cannot even be evolved across the point where $\alpha$ vanishes.  A possible counterargument to this reasoning is that the system enters the strong coupling regime whenever $\alpha \rightarrow 0$. We will argue, however, that the described singular behavior of the equation of motion persists in the same way in the non-linear theory: Consider the non-linear term $\gamma \phi \ddot \phi$ that will become important once $\alpha \sim \phi \gamma$.  In fact, this is the only relevant non-linear contribution, since any other term containing two time derivatives but more fields, such as $\phi^2 \ddot \phi$, will be subdominant due to the fact that $\phi$ itself is small, as explained. Thus the combination $(\alpha + \gamma \phi) \ddot{\phi}$ will determine the time evolution of the system, with the equation of motion
\begin{equation} \label{eom_stability}
(\alpha + \gamma \phi) \ddot{\phi} + \epsilon \dot \phi + \beta \phi = 0 ,
\end{equation}
Again, as long as $(\alpha + \gamma \phi) > 0$, the effective spring constant of the system grows large and confines $\phi$ to small values. At best, $\gamma \phi$ might have some positive value, so that $\alpha$ can become negative, but now $\alpha$ eventually drops to large negative values and will certainly overshoot the contribution $\gamma\phi$ which is still small due to the small $\phi$ fluctuations. Hence, even the sum $\alpha + \gamma \phi$ will pass through zero and result in a singularity of the system.

Let us elaborate a little bit more on the question why a vanishing coefficient $\alpha + \gamma \phi$ in front of the $\ddot \phi$ term entails an unacceptable singularity. We will name the time of zero crossing $t_0$, that is 
\begin{equation} \label{zero_crossing_ddot}
\alpha(t_0) + \gamma(t_0) \phi(t_0) = 0 .
\end{equation}
Assuming that $\ddot \phi$ is regular at $t_0$ yields the constraint $\epsilon(t_0) \dot \phi(t_0) +  \beta(t_0) \phi(t_0) = 0$ by virtue of the equation of motion (\ref{eom_stability}). Moreover, (\ref{zero_crossing_ddot}) yields the additional constraint $\phi(t_0) = - \alpha(t_0)/\gamma(t_0)$. These constraints completely spoil the Cauchy problem as they allow only one particular choice of initial conditions. This clearly illustrates the singular behavior of (\ref{eom_stability}) under the assumption of regular $\ddot \phi$. \newline
Thus, we try to abandon the assumption of regularity of $\ddot \phi$, and instead assume that $\ddot \phi \sim (\alpha + \gamma \phi)^{-1}$ around $t_0$. Taylor expansion of the vanishing coefficient gives the leading behavior  $\ddot \phi \sim (t-t_0)^{-\delta}$. The case $\delta = 2$ results in $\phi \sim \ln( | t-t_0 | )$ which is singular at $t=t_0$ and thus unacceptable. The same is true for $\delta > 2$, for which we obtain $\phi \sim (t-t_0)^{-\delta + 2 }$. If instead we have $\delta = 1$, $\phi$ would behave as $\phi \sim (t-t_0) \ln(| t-t_0 |) - (t - t_0 ) $, which would be well-defined at $t = t_0$. The term $\epsilon \dot \phi$ in (\ref{eom_stability}), however, would still be singular for this behavior of $\phi$, such that this behavior cannot give a solution to the equation (\ref{eom_stability}).

\subsection{Unitarity Bound} \label{sec_unitarity_bound}

At the level of the action for the SO(3) scalar $A$, 
the sign of the prefactor in front of the $\dot{A}^2$ term	
is crucial for the absence of negative norm states. (See 
\cite{Berkhahn:2010hc} for details.)
At the level of the equation of motion, this sign is determined
by the prefactor of the $\ddot{A}$ term which can be derived from combining equation (\ref{box_eom_A}) with 
the corresponding prefactor in the $\dot{E}$ term from (\ref{box_E}).
Combining both prefactors gives
\begin{equation}
	\frac{m^2 -  H^2 - \dot{H}}
	{(1-c_s^{\; 2})^2 (1 + 3 c_s^{\; 2})} \ddot{A} .
\end{equation}
Evidently, in the scalar sector unitarity seems to require 
that $m^2>H^2+\dot{H}$, which is precisely the bound 
(\ref{unitarity_bound}) found in \cite{Berkhahn:2010hc}
by employing Goldstone boson equivalence. 
As an important result, we re-derived this unitarity 
bound in a full-fledged cosmological perturbation analysis,
with a very important qualification: we find that the unitarity 
bound applies at all energies, and not just in the 
high-energy regime considered in \cite{Berkhahn:2010hc}.

In the following we solve the coupled 
equations of motions (\ref{box_eom_B}, \ref{box_eom_A}) for the 
scalars $A,B$ numerically, and analyze the stability of these 
solutions. For clarity, we subdivide the kinematical domain
in three subdomains: extreme sub-Hubble scales 
(${\bf k}^2/a^2\gg m^2, H^2$), intermediate scales, and
extreme super-Hubble scales (${\bf k}^2/a^2\ll m^2, H^2$).

\subsection{Extreme sub-Hubble scales} \label{sec_sub_hubble}

This regime has been investigated previously \cite{Berkhahn:2010hc}
employing the Goldstone boson equivalence as a diagnostic tool to
extract the leading short-distance dynamics. 

From the full, coupled set of linear differential equations (\ref{box_eom_A}-\ref{box_E}) these dynamics can be recovered by means of the adiabatic ansatz $A,B,E\propto e^{\mu t}$, which is best for large $\mathbf{k}_\mathrm{phys}$: Introducing the ansatz into the system of equations and solving the (biquadratic) secular equation $c_4\lambda^4+c_2\lambda^2+c_0=0$, which results to leading order in large $\mathbf{k}_\mathrm{phys}$, yields $\sqrt{2}\lambda=\pm\sqrt{(-c_2\pm\sqrt{c_2^2-4c_0c_4})/c_4}$, where $\lambda^2=\mu^2|\mathbf{k}_\mathrm{phys}|^2$. (Upper and lower signs can be chosen independently, which leads to four combinations.) In order to have a stable system, none of the eigenvalues may have a positive real part. Therefore, the presence of the outer $\pm$ implies that all eigenvalues must be purely imaginary. That necessitates that $c_2^2\le4c_0c_4$ and that $c_0$, $c_2$, and $c_4$ must have the same sign. Unitarity requires further that $c_4=2(H^2 + \dot H - m^2)$ is negative, reproducing Eq.~(\ref{unitarity_bound}). Hence, the system is stable when all coefficients are negative and $c_2^2\le4c_0c_4$. Then, from $c_0=(H^2 + \dot H/3 - m^2)w$ we reproduce Eq.~(\ref{stability_bound}) for $w>0$. 

For $w<0$ this relation would be exactly the other way round, implying that the system would never be stable. This phenomenon is known already from {\sl unmodified} general relativity \cite{Fabris:1996ua}, where a system filled by a perfect fluid with $w<0$ is always unstable as long as $\mathbf{k}_\mathrm{phys}$ is not very small. As it is already present in general relativity, this instability cannot have anything to do with the degree of freedom used in the Goldstone boson equivalence analysis, which is absent in general relativity. This explains why said instability goes unnoticed in this case. It is important to notice that in this respect a scalar field does not correspond to a perfect fluid \cite{Caldwell:1997ii}, which explains why this bound is also not obtained in \cite{Grisa:2009yy}.

Coming back to $w>0$, the bound derived from $c_2$ is always weaker than the stability bound (\ref{stability_bound}), which follows from $c_0$, or the requirement $c_2^2\le4c_0c_4$. For $w>1/3$ the requirement that $c_2^2\le4c_0c_4$ would be stronger than the bound (\ref{stability_bound}). A numerical analysis in the regime where $c_2^2\le4c_0c_4$ shows, however, that there is no instability as in the case where (\ref{stability_bound}) is violated. While the latter leads to a clear exponential explosion forwards and backwards in time, the latter manifests itself in a beat with an amplitude of the envelope that grows relatively mildly backwards in time. Here the requirement $c_2^2\le4c_0c_4$ obtained in the framework of the adiabatic analysis does not seem to give a relevant bound. Also in the case where the condition $c_2<0$ is violated, numerically no instability can be detected.

\begin{figure}[!ht]
  \centering
    \includegraphics[width=\textwidth]{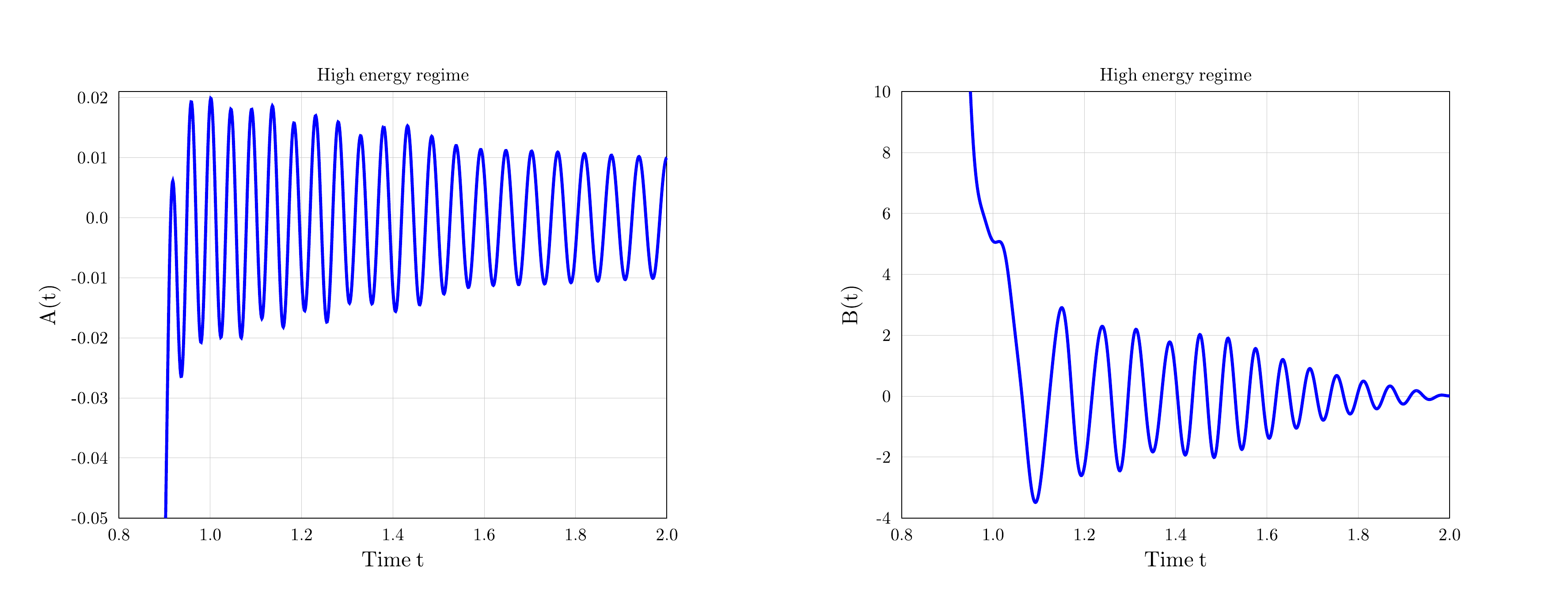}
  \caption{Scalars $A$ and $B$ during radiation domination deep inside the Hubble radius.}
  \label{fig_high_energy}
\end{figure}

Figure \ref{fig_high_energy} shows the numerical solution for 
the scalars $A$ and $B$ in a radiation-dominated universe ($c_s^{\; 2}=1/3$).
The parameters were chosen such that (units unspecified)
$k_{\rm phys}=250/\sqrt{t}$, $m=1/\sqrt{12}$, and $H=1/(2t)$. Hence, 
${\bf k}^2/a^2\gg m^2, H^2$ is guaranteed for times 
$t\in [0.8,2]$. The initial conditions have been chosen 
at $t=2$, such that the system is evolved backwards in time. 

Let us first investigate the behavior of $B$. For times $t>1$, $B$
is oscillating with a Hubble-damped amplitude, clearly showing 
a healthy hyperbolic evolution forward in time. 
Evolving backward in time, however, $B$ develops an instability 
for $t<1$. Indeed, the parameters have been chosen such that the 
stability bound (\ref{stability_bound}) is violated for $t<1$
This confirms the results of \cite{Berkhahn:2010hc}.
The behavior of $A$ is similar, except that it develops the instability
at an earlier cosmological time scale (which is later from the point of view 
of the system evolving backwards in time), and oscillates with a higher frequency as
compared to the scalar $B$. 

The basic properties of the solution are independent of the source's
equation of state in the interval $0\leq c_s^{\; 2} \leq 1$.
The case of a de Sitter source ($c_s^{\; 2}=-1$) is borderline, since
the parameter range for which the classical instability is triggered 
coincides precisely with the range of parameters for which unitarity gets violated.
Hence, the strong coupling regime goes hand in hand
with negative norm states. (See \cite{Berkhahn:2010hc} for details.)

\subsection{Intermediate scales} \label{sec_intermediate}

\begin{figure}[!ht]
  \centering
    \includegraphics[width=\textwidth]{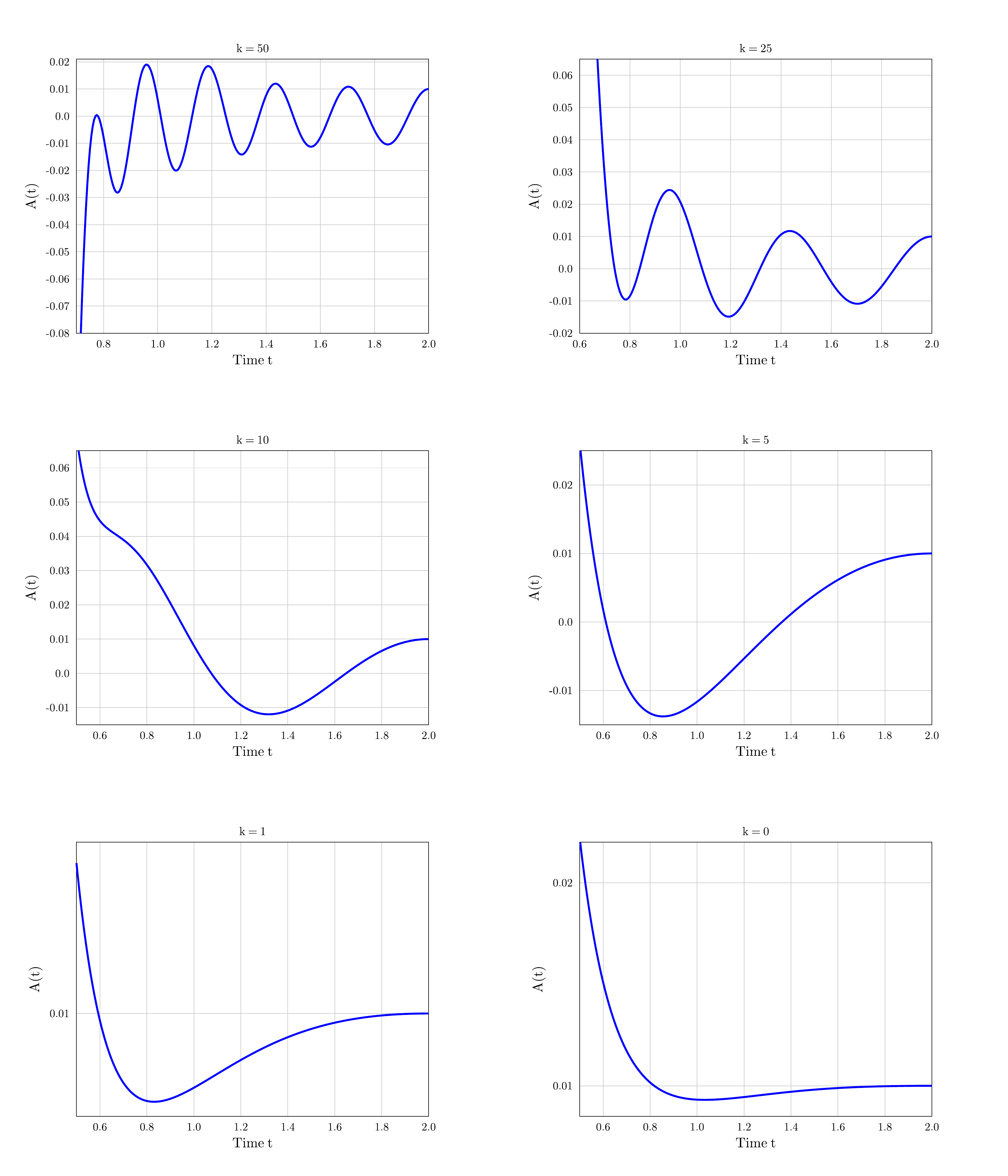}
  \caption{Numerical solution of $A(t)$ during radiation domination
		   for different values of $k_{\rm phys}(t=1)$.
 		   The other parameters have been chosen to be the same as for Figure 1.
			}
  \label{fig_low_energy_A}
\end{figure}

\begin{figure}[!ht]
  \centering
    \includegraphics[width=\textwidth]{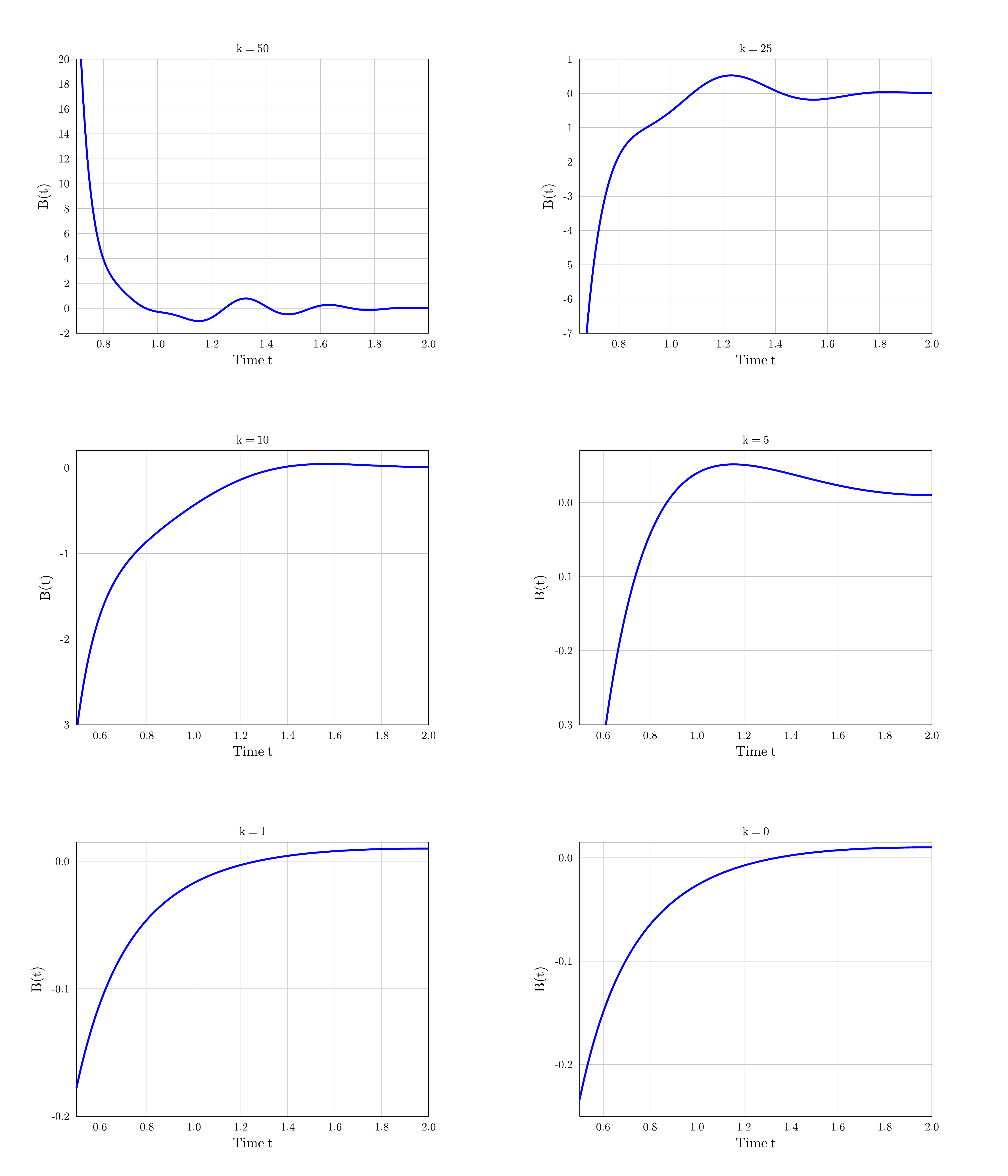}
  \caption{	Numerical solution for $B(t)$ during radiation domination
			   for different values of $k_{\rm phys}(t=1)$.
	 		   The other parameters have been chosen to be the same as in Figure 1. 
		   }
  \label{fig_low_energy_B}
\end{figure}

Figures \ref{fig_low_energy_A} and \ref{fig_low_energy_B} show the 
solutions for the scalars $A$ and $B$ during radiation domination
from intermediate to extreme super-Hubble scales, that is, for different 
values of the comoving wavenumber $k$ or, equivalently, 
for the physical wavenumber $k/a(t)$ at time $t=1$.
For convenience and clarity, the other parameters have been chosen precisely as 
in the previous section. 
Like in the previous case, the initial conditions have been chosen at $t=2$ and 
the scalar modes have been evolved backwards in time. For concreteness,
the initial conditions are given by $A=0.01$, $B=0.01$,
and $dA/dt=0$, $dB/dt=0$ at $t=2$. Note that the qualitative behavior 
of this dynamical system is quite insensitive to the choice of initial conditions,
in particular, with respect to the stability analysis.  

It can be seen that the scalar modes' behavior on intermediate scales (and also on extreme
super-Hubble scales, see next section) is very different from the dynamics in a 
normal neighborhood (see previous section). Compared to the latter case, 
the instability triggered at $t=1$ becomes less and less pronounced with decreasing
wavenumber. In order to appreciate this fact, notice the different ranges of 
mode amplitudes covered on the $y$-axes in figures \ref{fig_low_energy_A} and 
\ref{fig_low_energy_B} as compared to figure \ref{fig_high_energy}.
In fact, scalar fluctuations on super-Hubble scales show a power law behavior
which is triggered by the cosmological singularity (i.e. by the singular coefficients $H \propto 1/t$ etc), and which is clearly distinct from an instability
triggered by a non-hyperbolic evolution.

\subsection{Extreme super-Hubble scales} \label{sec_super_hubble}
In order to elucidate further this result, let us analyze the stability
of the scalar zero modes, which can be performed analytically. 
The zero modes of $A$ and $B$ satisfy (\ref{box_eom_A}-\ref{box_E}),
\begin{eqnarray}
	\label{box_eom_A0}
	\ddot{A}
	&=&
	- 3(1-w) H \dot{A} 
	- \left[2 m^2 - 6w\left(H^2-m^2/2\right)\right]A+
	\nonumber \\
	&& + w H \dot{\widetilde{B}}
	   + 2w\left(H^2-m^2/2\right) \widetilde{B}+
	\nonumber\\
	&& + H \dot{E}
	-m^2 E(A,B)
	\; ,\nonumber \\
	\\
	\label{box_eom_B0}
	\ddot{\widetilde{B}}
	&=&
	- 7 H \dot{\widetilde{B}} - 4 \left(H^2+m^2/2\right) \widetilde{B} +
	\nonumber \\
	&& -12 H \dot{A} - 12 H^2 A +
	\nonumber \\
	&& + 12 H^2 E(A,B) 
	\; ,
\end{eqnarray}	
where $E$ is expressed in terms of $A$ and $B$ as follows,
\begin{eqnarray}	
	\label{box_E0}
	&& \left[\dot{H} + \left(2-3w\right) H^2 - m^2\right]E(A,B) 
	= 
	\nonumber \\
	&& - \left(w-1/3\right)H \dot{A} 
	- \left[\dot{H}+\left(1+6w\right) H^2-\left(2+3w\right) m^2\right] A +
	\nonumber \\
	&& - \left(w-1/3\right)H\dot{\widetilde{B}}
	-
	(1/3)\left[\dot{H}+\left(1+6w\right) H^2-\left(2+3w\right) m^2\right] \widetilde{B}
	\; .
\end{eqnarray}
As a consequence, in this limit, the system of two coupled differential equations for $A$ and $B$
reduces to a single equation of motion for the linear combination $S\equiv A + \tilde B/3$,
\begin{equation}
	\left[
	C_2(w;t)\partial_t^{\; 2}
	+ 
	C_1(w;t)\partial_t
	+
	C_0(w;t)
	\right] S 
	= 0 \;,
\end{equation}	
where the coefficients $C_{2,1,0}$ depend on the equation of
state parameter $w$ of the source and on time via the Friedman background
evolution. Explicit expressions for these coefficients can be found in the appendix.

\begin{figure}[t]
  \centering
    \includegraphics[width=0.7\textwidth]{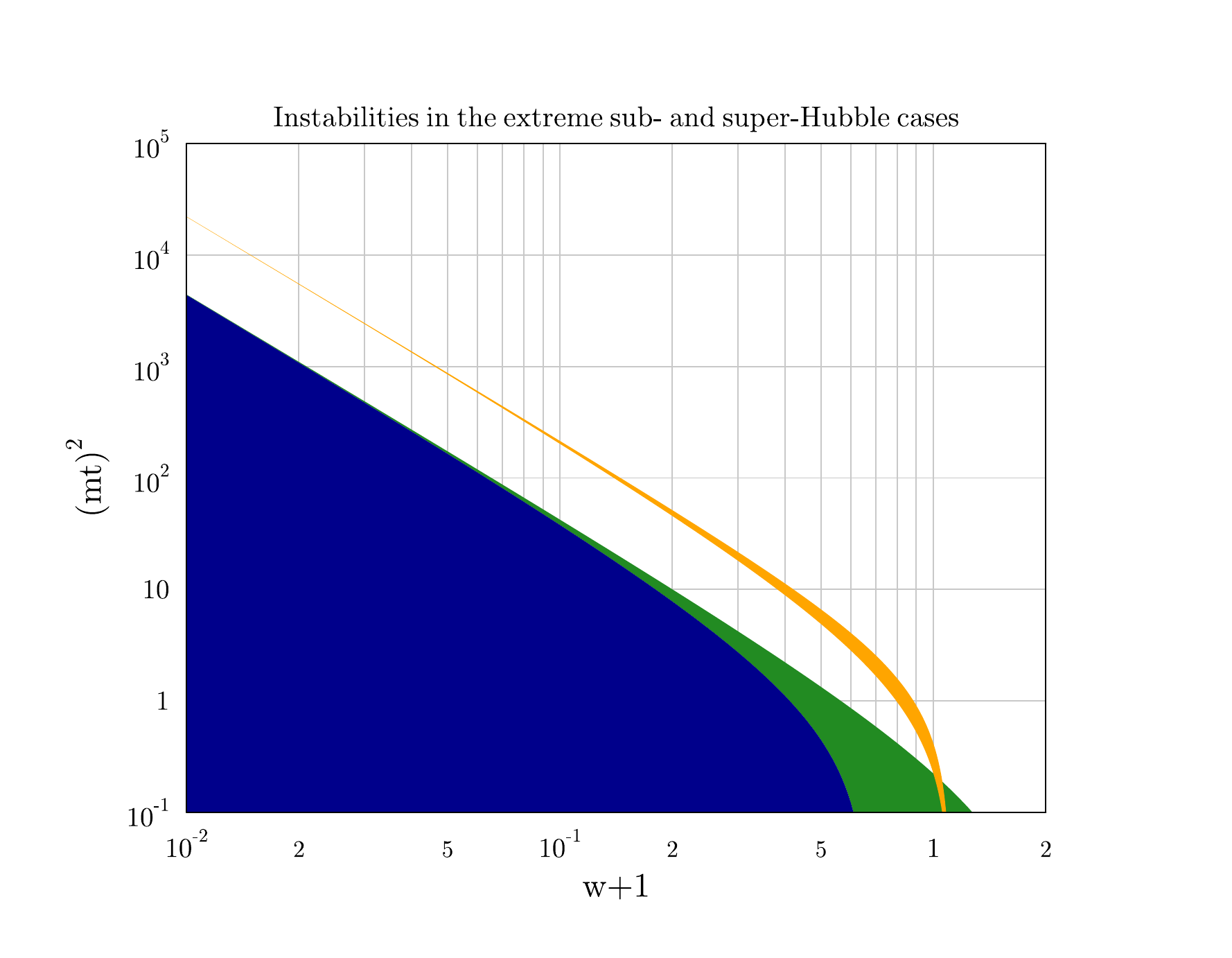}
  \caption{Instabilities in the extreme sub- and super-Hubble cases. In the orange region (top, detached), the system is classically unstable for $\mathbf{k}_\mathrm{phys}=\mathbf{0}$. The dark-blue region (bottom, left) depicts the region, where unitarity would be violated. In the green region (adjacent to the former), the system is classically unstable for large $\mathbf{k}_\mathrm{phys}$.}
  \label{fig_stability_k0}
 \end{figure}

A sufficient condition for hyperbolic evolution on the entire
Friedman manifold and thus, for classical stability,
is given by $C_1/C_2 > 0$ and $C_0/C_2>0$ for all times, for a given
source equation of state parameter $w$. We can analyze how these 
stability conditions depend on the parameter $w$ and time $t$.
The result is shown in Fig.~\ref{fig_stability_k0}, where the 
orange region corresponds to the classical instability region
for the zero mode $S$, and inside the dark-blue region
unitarity would be violated.
Figure \ref{fig_stability_k0} shows that for a source with equation
of state parameter $w \gtrapprox 0.11$, the zero mode's dynamics
is always stable, confirming our explicit numerical result 
for a radiation dominated Friedman universe discussed
in the previous section. For smaller values of $w$, when evolved backwards in time, the zero mode will always first enter the region of 
classical instability (orange), which signals the breakdown of perturbation theory. Evidently, it cannot enter the
unitarity violating region (dark-blue), without passing through
the strong coupling regime (orange).  
For large momenta, the area of classical instability moves downwards and comes to rest exactly on top of the area where unitarity would be violated, which, thus, still cannot be reached without first crossing the former (green).
Hence, in this sense the strong coupling
regime self-protects the scalar zero mode from unitarity violation, as well. 
As a consequence, it is not clear at all whether the thus diagnosed unitarity violating region 
is of physical relevance, as it lies well outside the perturbative regime. 
We can turn this argument around and conclude that no inconsistency
is present within the perturbative regime.  
%


\section{Conclusion}

In summary, using cosmological perturbation theory,
we have proven the consistency of the most relevant
Einstein--Hilbert deformation in the perturbative regime. 
The deformation itself achieves consistency via a self-protection
mechanism that pushes potential unitarity violations
beyond the weak coupling regime. 
This confirms previous studies concerning the deformation's
nontrivial stability dynamics, based on a Stückelberg completion
of the deformation in conjunction with the Goldstone boson
equivalence \cite{Berkhahn:2010hc}. 
Most importantly, this work extends the self-protection mechanism to
encompass the entire kinematical domain, ranging from
sub- to super-Hubble scales. 

It would be interesting to study the proposed non-linear theories \cite{deRham:2010ik, deRham:2010kj, Hassan:2011vm} with a rigid FRW background to see whether they non-linearly exhibit the self-protection mechanism.

As discussed in great detail in \cite{Berkhahn:2011jh}, the self-protection phenomenon is a prime example for the recently conceived classicalization mechanism \cite{Dvali:2011th,Dvali:2011nj,Dvali:2010ns,Dvali:2010jz,Dvali:2010bf} and extends it further to free ﬁeld dynamics on curved backgrounds. 

\section{Appendix}

\subsection{Derivation of the evolution equations in the scalar sector}

To start with, consider the part of the momentum conservation equation $\delta \nabla^{\mu} T_{\mu i} = 0$ that is built up from a derivative $\partial_j $ of a scalar variable:
\begin{equation} \label{momentum_partial}
\partial_j \left[ \delta p + \partial_0 \left( \left( \bar{\rho} + \bar{p} \right) \delta u \right) + 3 H  \left( \bar{\rho} + \bar{p} \right) \delta u + \frac{1}{2} \left( \bar{\rho} + \bar{p} \right) E \right] = 0
\end{equation}

We will specialize to an equation of state of the simple form $\delta p = \frac{\partial p}{\partial \rho} \delta \rho $. By doing so, we restrict ourselves to the case of a one-component system. The more complicated case of multi-component systems can be investigated, but one needs further special information about the system (for example the separate energy-momentum conservation of each component if they do not interchange energy and momentum). Further, using the Friedmann equations, one easily shows that $8 \pi G  \left( \bar{\rho} + \bar{p} \right) = -2 \dot{H} $. The fluctuation $\delta u $ can be expressed in terms of metric variables using the $i0 $-equations of (\ref{eom2}), where one again extracts the contributions built from a derivative of scalar variables,
\begin{equation} \label{delta_u}
8 \pi G \left( \bar{\rho} + \bar{p} \right) \partial_j \delta u = \partial_j \left[ -H E + \dot{A} - m^2 (aF) \right] .
\end{equation}
Using this in Eq.~(\ref{momentum_partial}), together with equation (\ref{effgaugeS2}), one derives
\begin{equation} \label{partial_eom_A}
\partial_j \left[ 8\pi G \frac{\partial p}{\partial \rho} \delta \rho - H \dot{E} - \left( 3 H^2 + 2 \dot{H} \right) E +m^2 E+ \ddot{A} + 3H \dot{A}  + 2 m^2 A \right] = 0 .
\end{equation}
Since the spatial divergence of the bracket in (\ref{partial_eom_A}) vanishes identically, we know that the expression in the bracket is equal to some function of time alone. As we know from the basic equation (\ref{eom2}) that $h_{\mu \nu} = 0$, $T_{\mu \nu} = 0 $ (which corresponds to $A=0$, $B=0$, $E=0$, $\delta \rho=0$, etc.) must be a solution, this function of time must be identically zero. Hence, we obtain
\begin{equation} \label{eom_A}
8 \pi G \frac{\partial p}{\partial \rho} \delta \rho - H \dot{E} - \left( 3 H^2 + 2 \dot{H} \right) E +m^2 E  + \ddot{A} + 3H \dot{A}  + 2 m^2 A = 0 .
\end{equation}

Next, we will consider the $ij $-equations of (\ref{eom2}) from which we extract the part of the form $\partial_i \partial_j S $ with $S$ a scalar. This gives
\begin{equation} \label{ij_derivative_scalar}
\partial_i \partial_j \left[ E + A - a^2 \ddot{B} - 3 a \dot{a} \dot{B} - 2 m^2 a^2 B + 2 a \dot{F} + 4 \dot{a} F  \right] = 0 .
\end{equation}
Using (\ref{effgaugeS2}) we can reexpress
\begin{equation}
\partial_j \left( 2 a \dot{F} + 4 \dot{a} F \right)  = \partial_j \left( 2 (aF\dot{)} + 2 \dot{a} F \right) = \\ = \partial_j \left( - 4 \dot{a} F - 2E - 4A \right) .
\end{equation}
Inserting this in (\ref{ij_derivative_scalar}) and taking the trace of the result gives
\begin{equation}
 -  \ddot{\tilde{B}} - 3 H \dot{\tilde{B}} - 2 m^2 B  - 4 H \frac{\Delta}{a} F - \frac{\Delta}{a^2}E - 3 \frac{\Delta}{a^2} A  = 0 .
\end{equation}
Finally, using (\ref{effgaugeS1}) we obtain,
\begin{equation} \label{eom_B}
-  \ddot{\tilde{B}} - 7 H \dot{\tilde{B}} - 4 H^2 \tilde{B} - 2 m^2 B    - 12 H \dot{A} - 3 \frac{\Delta}{a^2} A - 12 H^2 A   - \frac{\Delta}{a^2}E+ 12 H^2 E  = 0 .
\end{equation}
This equation is the first of the two basic evolution equations in the scalar sector, see (\ref{box_eom_B}).

The $00$-equation of (\ref{eom2}) gives
\begin{eqnarray} \label{00_scalar}
-4 \pi G \left( 1 + 3 \frac{\partial p}{\partial \rho} \right) \delta \rho  = &- \frac{3}{2} H \dot{E} - \frac{\Delta}{2a^2} E  -3 \left(H^2 + \dot{H} \right) E + \frac{3}{2} m^2 E +  \nonumber \\
&   + \frac{3}{2} \ddot{A} + 3 H \dot{A} + \frac{3}{2} m^2 A +\nonumber \\
 & + \frac{1}{2} \ddot{\tilde{B}} + H \dot{\tilde{B}} +  \frac{1}{2} m^2 \tilde{B}  \nonumber \\
 &    - \frac{1}{a^2} (a\Delta F\dot{)} .
\end{eqnarray}
Using (\ref{effgaugeS1}) one can eliminate $F $ from (\ref{00_scalar}),
\begin{eqnarray} \label{dr1}
-4 \pi G \left( 1 + 3 \frac{\partial p}{\partial \rho} \right) \delta \rho  = & + \frac{3}{2} H \dot{E} - \frac{\Delta}{2a^2} E +  3 H^2 E + \frac{3}{2} m^2 E \nonumber \\
 & - \frac{3}{2} \ddot{A} - 6 H \dot{A} - 3 \dot{H} A  - 6 H^2 A + \frac{3}{2} m^2 A \nonumber \\
 &   - \frac{1}{2} \ddot{\tilde{B}}  - 2 H \dot{\tilde{B}} - \dot{H} \tilde{B} - 2 H^2 \tilde{B} + \frac{1}{2} m^2 \tilde{B} .
\end{eqnarray}

The $jk $-equations proportional to $\delta_{jk} $ give
\begin{eqnarray} 
-4 \pi G \left(1 - \frac{\partial p}{\partial \rho} \right) \delta \rho = & \frac{1}{2} H \dot{E} + \left( 3H^2 + \dot{H} \right) E - \frac{1}{4} m^2 E \nonumber \\ 
& - \frac{1}{2} \ddot{A} + \frac{\Delta}{2a^2} A  - 3H \dot{A} - \frac{5}{4} m^2 A + \nonumber \\
& -\frac{1}{2} H \dot{\tilde{B}}   - \frac{1}{4} m^2 \tilde{B} \nonumber \\
& + H \frac{\Delta}{a} F  .
\end{eqnarray}
Let us again eliminate $F$ using Eq.~(\ref{eom2}),
\begin{eqnarray} \label{dr2}
-4 \pi G \left(1 - \frac{\partial p}{\partial \rho} \right) \delta \rho = & \frac{1}{2} H \dot{E} + \dot{H} E - \frac{1}{2} m^2 E  \nonumber \\
& - \frac{1}{2} \ddot{A} + \frac{\Delta}{2a^2} A   + 3 H^2 A +  - \frac{5}{2} m^2 A + \nonumber \\
& + \frac{1}{2} H \dot{\tilde{B}} + H^2 \tilde{B}  - \frac{1}{2} m^2 \tilde{B} .
\end{eqnarray}
Inserting this expression for $\delta \rho$ into (\ref{eom_A}) results in a second independent evolution equation in the scalar sector (\ref{box_eom_A}).

Equating (\ref{dr1}) and (\ref{dr2}) allows us to eliminate $\delta \rho $
\begin{eqnarray} \label{dreq}
\frac{1}{1-\frac{\partial p}{\partial \rho}}  \biggl[  \dot{H} E + \frac{1}{2} H \dot{E} - \frac{1}{2} m^2 E  - \frac{1}{2} \ddot{A} + \frac{\Delta}{2a^2} A + 3 H^2 A - \frac{5}{2} m^2 A + \frac{1}{2} H \dot{\tilde{B}}  +  H^2 \tilde{B}   +  \nonumber \\
 - \frac{1}{2} m^2 \tilde{B} \biggr] = \frac{1}{1+3 \frac{\partial p}{\partial \rho}} \biggl[ \frac{3}{2} H \dot{E} -  3 H^2 E + \frac{3}{2} m^2 E - \frac{3}{2} \ddot{A} +  \frac{3\Delta}{2a^2} A - 3 \dot{H} A  + \frac{3}{2} m^2 A   \nonumber \\
 + \frac{3}{2} H \dot{\tilde{B}}  - \dot{H} \tilde{B}  +  \frac{3}{2} m^2 \tilde{B} \biggr] .
\end{eqnarray}
Here, in addition we have used Eq.~(\ref{eom_B}) to eliminate $\ddot{B}$.

Our ultimate aim is to express $E$ in terms of $A $ and $B $. In the first place, one might think that Eq.~(\ref{eom_B}) does the job for every mode $\vec{k}_{phys} $, but the problem with this equation is that the resulting expression for $E $ would contain $\ddot{ \tilde{B} } $, so that whenever $\dot{E} $ appears one would get three time derivates on $\tilde{B} $. This is something we should, if possible, try to avoid for the sake of tractability, and indeed, this is possible. One way (among others) is first to derive an additional equation in $A$, $B$, and $E $ by just using the constraints (\ref{effgaugeS1}) and (\ref{effgaugeS2}):
\begin{eqnarray}
& \Bigl( \frac{\Delta}{a}F \dot{\Bigr)} & =  \Bigl( -3 H E + 3 H A + H \tilde{B} + 3 \dot{A} + \dot{\tilde{B}} \dot{\Bigr)} =  \nonumber \\ 
= & \Bigl( \frac{1}{a^2} \Delta (aF) \dot{\Bigr)} & =  \frac{1}{a^2} \Delta (aF \dot{)} - 2 H \frac{\Delta}{a} F = - 5 H \frac{\Delta}{a} F - \frac{\Delta}{a^2} E - 2 \frac{\Delta}{a^2} A = \nonumber \\
& & = 15 H^2 E - 15 H^2 A - 5 H^2 \tilde{B} - 15 H \dot{A} - 5 H \dot{\tilde{B}} - \frac{\Delta}{a^2} E + \nonumber \\
& & \hspace{12pt} - 2 \frac{\Delta}{a^2} A .
\end{eqnarray}
Using in addition Eq.~(\ref{eom_B}) to eliminate $\ddot{B} $ this can be cast into the form
\begin{eqnarray} \label{eom_intermediate}
&3 \ddot{A} + 6 H \dot{A}  - \frac{\Delta}{a^2} A + 3 \dot{H} A + 3 H^2 A - 3 H \dot{E} - 3 \dot{H} E   + \nonumber \\
 &- H \dot{\tilde{B}} + \dot{H} \tilde{B} - 2 m^2 \tilde{B} + H^2 \tilde{B} - 3 H^2 E   = 0 .
\end{eqnarray}
As it happens, the ratio of the coefficients in front of $\dot{E} $ and $\ddot{A} $ coincides for equations (\ref{dreq}) and (\ref{eom_intermediate}). Therefore, by appropriately adding both equations, one eliminates $\dot{E} $ and $\ddot{A} $ at once, leaving an equation, which can be solved explicitly for $E $ in terms of $A $ and $B$ and their first derivatives. This equation is given by Eq.~(\ref{box_E}).

\subsection{The evolution equations for $\bf k = 0$}
\label{sec_app_k0}
In the case $\vec{k}_{phys} = 0$ the equations of motion (\ref{box_eom_A}, \ref{box_eom_B}, \ref{box_E}) reduce to
\begin{align}  
& E =  
\frac{\left[\dot{H}+H^2\left(1+6w\right)-\frac{m^2}{2}\left(2+3w\right)\right]\left(A+\frac{1}{3}\tilde{B}\right)+
H\left(\dot{A}+\frac{1}{3}\dot{\tilde{B}}\right)\left(-1+3w\right)}
{-\dot{H}- H^2 \left(2-3w\right)  + \frac{m^2}{2}}
\end{align}
\begin{equation} 
12H^2E - 12 H^2 \left(A+\frac{1}{3}\tilde{B}\right) - 12 H \left(\dot{A}+\frac{1}{3}\dot{\tilde{B}}\right) -  \ddot{\tilde{B}} - 3 H \dot{\tilde{B}} - m^2 \tilde{B} = 0
\end{equation}
\begin{align}  \label{eom_C_k0}
\left[H^2(4-6w)+m^2\left(1+\frac{3}{2}w\right)\right] \left(A+\frac{1}{3}\tilde{B}\right)     
+ H (7-3w) \left(\dot{A}+\frac{1}{3}\dot{\tilde{B}}\right)
+\nonumber \\
+ \left(H^2(-7+3w) - 2 \dot{H}+\frac{m^2}{2} \right) E - H \dot{E} + \left(\ddot{A}+\frac{1}{3}\ddot{\tilde{B}}\right) = 0
\end{align}
i.e.~the equation of motion for $S \equiv A + \frac{1}{3} \tilde{B}$ (\ref{eom_C_k0}) decouples, which we will abbreviate by
\begin{align}  \label{eom_C_k0_abb}
C_2(t) \ddot S
+C_1(t) \dot S
+C_0(t) S = 0
\end{align}
with $C_0(t), C_1(t)$ and $C_2(t)$ given by
\begin{eqnarray}
C_0  &= &\nonumber\\
&=&
\left[H^2(4-6w)+m^2\left(1+\frac{3}{2}w\right)\right]
+\nonumber\\
&&+
\left[H^2(-7+3w)-2\dot{H}+\frac{m^2}{2}\right]
\left[\dot{H}+H^2\left(1+6w\right)-\frac{m^2}{2}\left(2+3w\right)\right]\left[-\dot{H}-H^2(2-3w)+\frac{m^2}{2}\right]^{-1}
-\nonumber\\
&&-
H\left\{
\left[-\dot{H}-H^2(2-3w)+\frac{m^2}{2}\right]^{-1} \left[\ddot{H}+2\dot{H}H(1+6w)\right]
\right.-\nonumber
\\
&&\left.
\left[-\ddot{H}- 2\dot{H}H(2-3w)\right]\left[\dot{H}+H^2(1+6w)-\frac{m^2}{2}(2+3w)\right]\right\}
\\
C_1&=&
(7-3w)H
+\nonumber\\
&&+
H\left(H^2(-7+3w) - 2 \dot{H}+\frac{m^2}{2} \right)
\left(-1+3w\right)
[-\dot{H}- H^2 \left(2-3w\right)  + \frac{m^2}{2}]^{-1}
-\nonumber\\
&&-H
\left\{\dot{H}+H^2\left(1+6w\right)-\frac{m^2}{2}\left(2+3w\right)\right\}\left[-\dot{H}- H^2 \left(2-3w\right)  + \frac{m^2}{2}\right]^{-1}
\nonumber\\
&&
-H\left\{
\dot{H}\left(-1+3w\right)(-\dot{H}- H^2 \left(2-3w\right)  + \frac{m^2}{2})^{-1}
-H\left(-1+3w\right)(-\ddot{H}- 2\dot{H}H \left(2-3w\right))\right\}
\\
C_2&=&
1-\frac{H^2\left(-1+3w\right)}
{-\dot{H}- H^2 \left(2-3w\right)  + \frac{m^2}{2}}
=
\frac{-\dot{H}- H^2  + \frac{m^2}{2}}
{-\dot{H}- H^2 \left(2-3w\right)  + \frac{m^2}{2}}
\end{eqnarray}


\section*{Acknowledgments}

The authors would like to thank 
Gia Dvali,
Fawad Hassan, 
Justin Khoury, 
Michael Kopp, 
Florian Kühnel,
Parvin Moyassari,
Slava Mukhanov, 
and 
Florian Niedermann 
for helpful and inspiring discussions.
The authors acknowledge gratefully the hospitality of the Nordic Institute for Theoretical Physics.
DDD acknowledges gratefully the hospitality of the Arnold Sommerfeld Center and the Excellence Cluster Universe.
The work of SH was supported by the DFG cluster of excellence 'Origin and Structure of the Universe'
and by TRR 33 'The Dark Universe'.
The work of FB was supported by TRR 33 'The Dark Universe'.


\end{document}